\documentclass[sigconf]{acmart}

\usepackage{subcaption}
\usepackage{booktabs}
\usepackage{tabularx}
\usepackage{array}
\AtBeginDocument{%
  }

\renewcommand\footnotetextcopyrightpermission[1]{} 
\fancypagestyle{plain}{
  \fancyhead{}
  \fancyfoot[C]{\footnotesize\thepage}
}

\begin{document}

\title{Scaling Laws For Data-Constrained Dense Retrieval}
\author{
Bowen Fu$^{2}$, Jingtao Zhan$^{1\dagger}$\\[4pt]
$^{1}$ Tsinghua  Shenzhen International Graduate School \\ $^{2}$ Xi'an Jiaotong University\\[2pt]
}



\begin{abstract}
Dense retrieval models are typically obtained by fine-tuning pretrained language models on supervised retrieval data, where high-quality annotations or interaction logs are often limited and reused across many epochs. This data-constrained regime differs from the abundant-data setting assumed by standard scaling laws, making it unclear how the best attainable retrieval performance changes with model size and unique fine-tuning data. We study this question through the minimum achievable evaluation loss (MAEL), defined as the lowest evaluation loss reached during training under fixed model-size and data-budget constraints. We first show that learning-rate choices strongly affect optimization trajectories but have little effect on the final attainable MAEL: after rescaling the optimization-step axis, learning curves collapse across learning rates, and CKA analysis shows that checkpoints reaching MAEL learn highly similar representations. We then fine-tune 35 BERT-family dense retrievers on MS MARCO subsets and find that MAEL is accurately described by a simple multiplicative power law over non-embedding parameters and unique samples, outperforming commonly used additive scaling formulations. We further derive a scaling law for the training step at which MAEL is reached and show that optimal weight decay follows an approximately invariant EMA timescale, yielding lower MAEL and stronger retrieval performance on both in-domain and out-of-domain benchmarks. Finally, combining loss and step scaling suggests that, in our observed regime, increasing pretrained model size is approximately 1.46 times as fine-tuning-compute-efficient as increasing unique in-domain data for reducing MAEL.
\end{abstract}


\maketitle

\renewcommand{\thefootnote}{\dag}
\footnotetext[2]{Corresponding author (jingtaozhan@tsinghua.edu.cn)}

\section{Introduction}
Large language model scaling laws have shown that increasing model size and training data consistently leads to improved performance~\cite{kaplan_scaling, chinchilla_scaling}. Similar observations have also been made in the information retrieval literature, where prior work demonstrates that scaling laws continue to hold for dense retrieval models~\cite{scaling_dr}. Unlike large language models, which are typically pretrained on massive amounts of unlabeled text, dense retrieval models are generally obtained by fine-tuning pretrained language models on downstream retrieval tasks~\cite{dpr, ance, rocket_qa, m3_embedding, qwen3_embedding}. However, high-quality supervised retrieval datasets, such as those constructed through manual annotation or collected from user interaction logs~\cite{nq, msmarco, orcas, trivia_qa}, are relatively scarce and are several orders of magnitude smaller than pretraining corpora. As a result, repeatedly reusing the same training data over multiple epochs has become standard practice in the fine-tuning paradigm~\cite{dpr, ance, rocket_qa}.

Existing scaling laws, however, have largely been developed under the compute-optimal regime, which assumes access to abundant, unique training data. Such assumptions no longer hold in many practical settings, such as dense retrieval. Motivated by the increasing scarcity of high-quality data, recent studies have investigated data-constrained training and proposed scaling laws that characterize the evolution of model performance throughout optimization~\cite{scaling_laws_and_interpretability_of_learning_from_repeated_data, scaling_data_constrained_lm, pretraining_under_infinite_compute, prescriptive_scaling_laws}. While these studies successfully model the training dynamics, practical deployment is often less concerned with the entire optimization trajectory than with the best performance achievable under a fixed model size and data budget. Consequently, despite recent progress in scaling under data constraints, a fundamental question remains unanswered: what is the best performance achievable with a fixed amount of unique training data and a fixed model size in information retrieval post-training?

In practice, the performance observed at a given training stage depends not only on model size and data, but also on the optimization procedure, making it difficult to disentangle the effect of scaling from that of the training recipe. We find that, despite following substantially different optimization trajectories under different learning rates, models with the same model size and unique-data budget reach nearly identical minimum evaluation losses. Moreover, their learning curves can be aligned through a simple rescaling of the optimization-step axis, suggesting that the learning-rate choice primarily affects the rate at which a given level of performance is reached rather than the attainable performance itself. This motivates using the minimum achievable evaluation loss (MAEL) reached during training as a stable, recipe-invariant measure of attainable performance. We therefore study how this quantity scales with model size and unique data under a fixed fine-tuning compute budget.

Building on this observation, we study 35 model-size and unique-data-budget configurations spanning a wide range of scales. We find that lower MAEL consistently corresponds to better downstream retrieval effectiveness and that MAEL is accurately described by a simple multiplicative power law. This scaling relationship enables us to quantify the marginal benefits of increasing model size and unique data, providing a principled framework for improving fine-tuning compute efficiency. We further investigate the effect of weight decay and find that appropriately chosen regularization can further reduce MAEL and improve retrieval performance; the optimal weight decay follows a scaling trend consistent with recipes previously observed in large-scale pretraining~\cite{how_to_set_wd}. Our main contributions are as follows:
\begin{itemize}
    \item We show that, under data-constrained post-training, learning-rate schedules substantially affect optimization trajectories but have little influence on the minimum achievable evaluation loss after convergence.
    \item We identify the minimum achievable evaluation loss as a useful characterization for fixed model size and unique training data, and demonstrate that it follows a simple multiplicative scaling law across 35 models.
    \item We derive a scaling law for the optimization step at which MAEL is reached and show that combining the loss and step scalings yields a clear compute-efficiency tradeoff between model size and unique data.
\end{itemize}

\section{Related Works}

\subsection{Dense retrieval}
Dense retrieval has become the dominant paradigm for neural information retrieval by representing queries and documents with dense embeddings learned by neural encoders. DPR~\cite{dpr} first demonstrated that dual-encoder models trained on supervised query--document pairs substantially outperform traditional sparse retrieval methods such as BM25~\cite{bm25}. Subsequent work improved retrieval quality through more effective hard-negative mining~\cite{ance, dr_with_hard_negatives}, stronger pretrained encoders~\cite{e5,bge,roberta,modern_bert,qwen3_embedding}, and knowledge distillation~\cite{tas_b, rocket_qa}.

Unlike language model pretraining, dense retrievers are typically fine-tuned from pretrained language encoders using supervised datasets containing only tens or hundreds of thousands of labeled query--document pairs. Consequently, models are commonly trained for dozens of epochs~\cite{dpr,ance}, repeatedly revisiting the same samples. This repeated-data training regime differs fundamentally from the continually expanding data setting assumed in most existing scaling-law studies and motivates the analyses presented in this work.

\subsection{Scaling laws}
Neural scaling laws describe predictable empirical relationships between model performance and scaling factors such as model size, training data, and compute. Early work showed that generalization error often follows power-law trends as data and model scale increase~\cite{hestness_scaling}. Kaplan et al.~\cite{kaplan_scaling} established such laws for autoregressive language models, showing that language-modeling loss scales smoothly with parameter count, dataset size, and training compute. Hoffmann et al.~\cite{chinchilla_scaling} later refined this view by studying compute-optimal training and showing that, under a fixed compute budget, model size and training tokens should be scaled in a more balanced way. These studies make scaling laws useful not only as descriptive models of performance, but also as practical tools for extrapolating large-scale behavior from smaller training runs.

Beyond language-model pretraining, scaling-law analyses have been extended to other architectures~\cite{scaling_vit, data_scaling_laws_in_nmt, constructive_rrediction_of_generalization_error, scaling_laws_for_recommendation_models}, modalities~\cite{scaling_vit, data_scaling_laws_in_nmt, constructive_rrediction_of_generalization_error}, and downstream applications~\cite{scaling_laws_for_autoregressive_generative_modeling, scaling_laws_for_recommendation_models}. More recently, scaling-law analyses have also been extended to fine-tuning, where limited data and repeated exposure to the same samples introduce optimization dynamics that differ substantially from the pretraining regime. Some work~\cite{scaling_laws_for_transfer, scaling_data_constrained_lm} modeled the diminishing information gained from repeated data passes during fine-tuning, while ~\cite{finetune_scaling_zhang_iclr24} derived scaling laws for different fine-tuning strategies, including full-parameter tuning, LoRA, and prompt tuning. Subsequent studies further investigated scaling behavior under domain shift and catastrophic forgetting~\cite{forgetting_scaling_arxiv, forgetting_scaling_icml25}.

Since dense retrieval is typically trained through supervised fine-tuning on pretrained language encoders, recent work has shown that dense retrieval also exhibits predictable scaling behavior with respect to model and data scales~\cite{scaling_dr}. However, existing dense retrieval scaling laws assume that each training sample is observed only once and therefore do not model the repeated-sample regime that is ubiquitous in practical fine-tuning. Consequently, they cannot characterize the overfitting dynamics or predict the minimum achievable evaluation loss under realistic fine-tuning settings.

\subsection{Data constrained training}
In recent years, increasing attention has been devoted to understanding the training dynamics of language models under data-constrained settings, leading to a series of scaling laws that characterize their learning behavior. Early studies introduced a small fraction of repeated samples into the training corpus and observed a double-descent phenomenon, showing that data repetition can degrade model performance~\cite{scaling_laws_and_interpretability_of_learning_from_repeated_data}. Subsequent work investigated the learning dynamics when the entire dataset is repeatedly reused, demonstrating that repeated data can provide learning efficiency comparable to fresh data during the early stages of reuse~\cite{scaling_data_constrained_lm}. Building on these findings, more recent studies have explored how to optimize pretraining under a fixed data budget with effectively unlimited compute~\cite{pretraining_under_infinite_compute}, as well as proposed improved scaling laws that accurately capture the U-shaped learning curves induced by overfitting on reused data~\cite{prescriptive_scaling_laws}.

However, existing studies primarily aim to model the entire learning trajectory, while paying relatively little attention to the scaling behavior of the minimum evaluation loss. Compared with the full learning curve, the minimum evaluation loss is substantially more stable across different hyperparameter settings, making it a more suitable target for scaling law analysis.

\section{Methodology}

\subsection{Minimum Achievable Evaluation Loss}

\begin{figure*}[t]
    \centering
    \begin{subfigure}{0.49\linewidth}
        \centering
        \includegraphics[width=\linewidth]{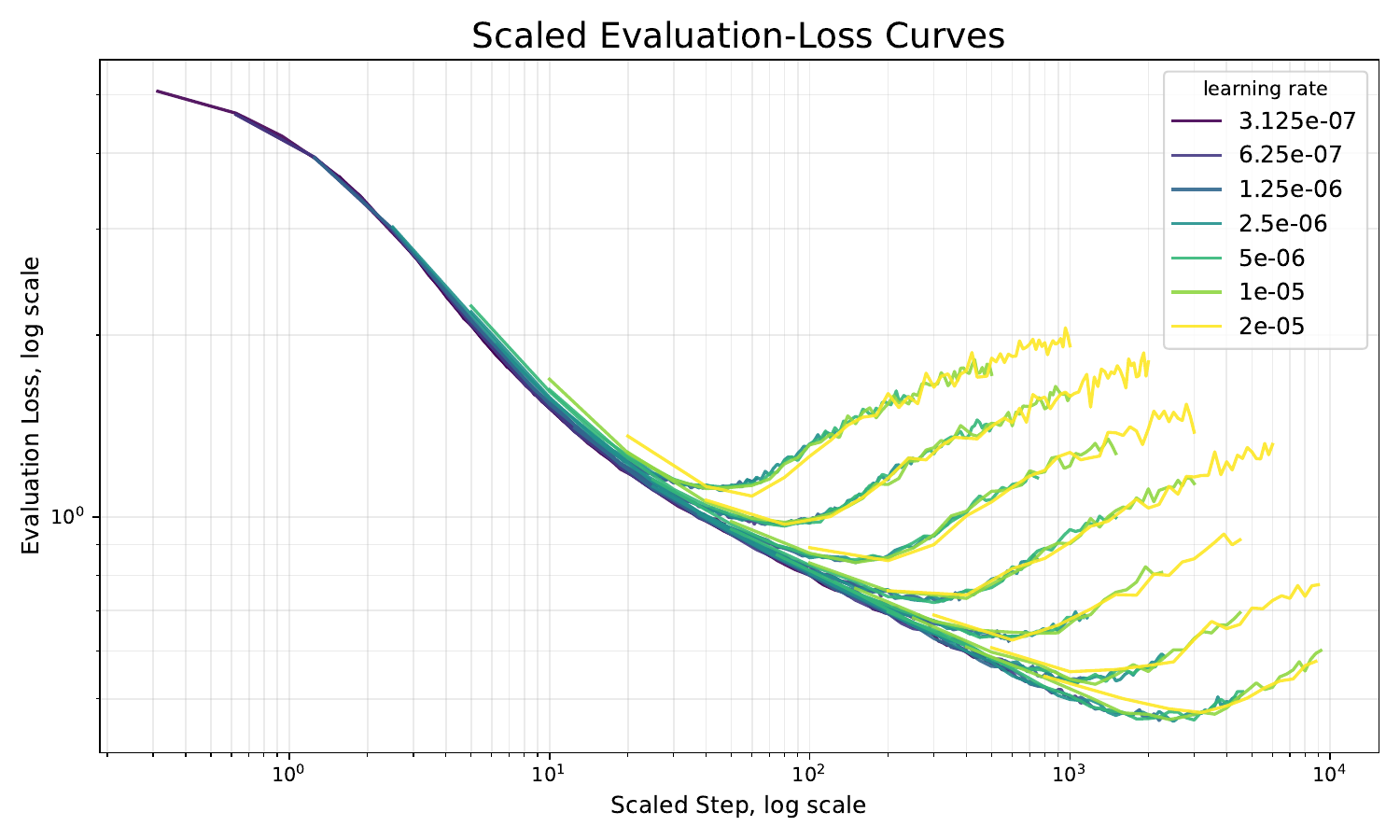}
        \label{fig:another}
    \end{subfigure}
    \begin{subfigure}{0.3\linewidth}
        \centering
        \includegraphics[width=\linewidth]{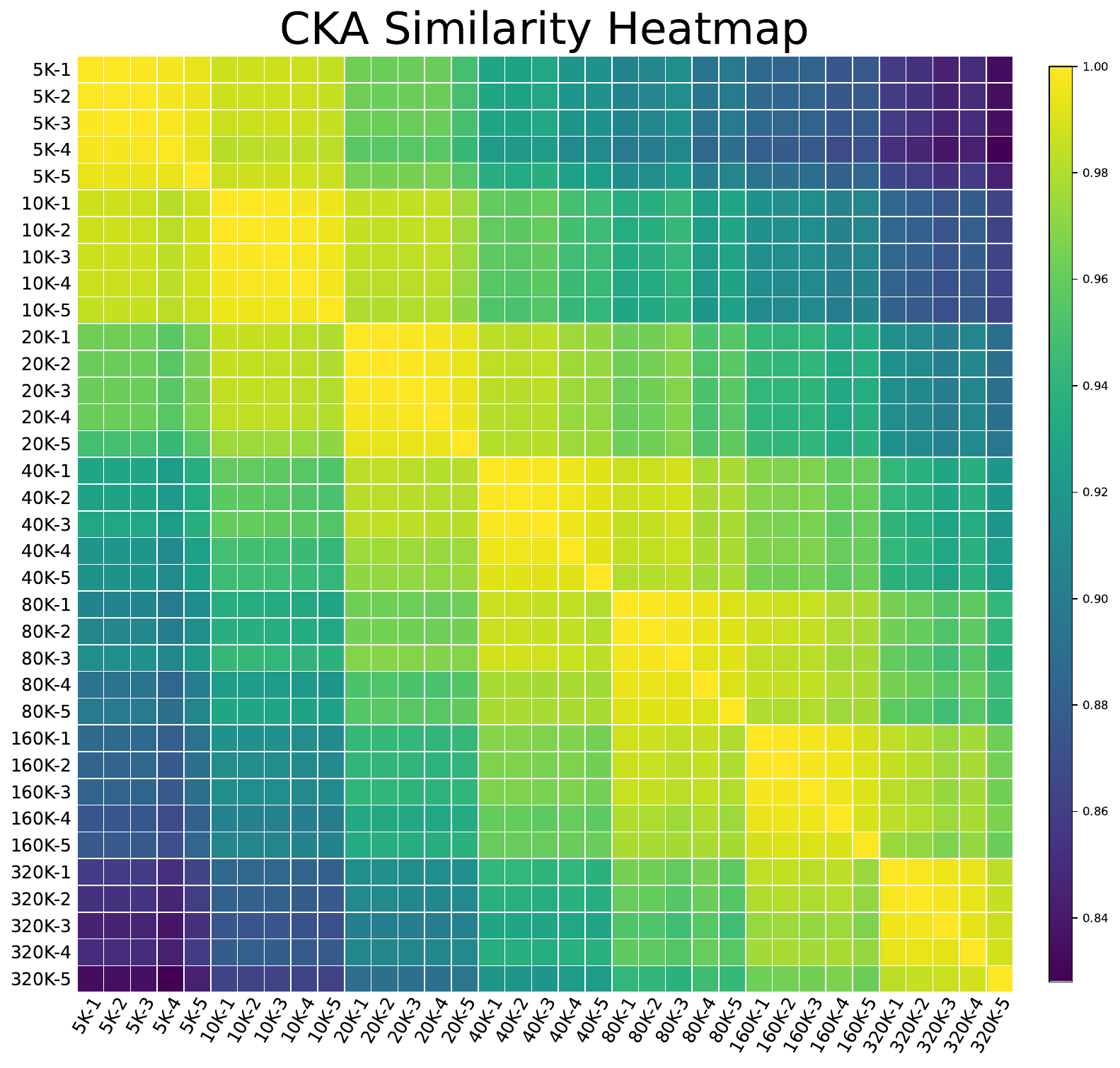}
        \label{fig:cka}
    \end{subfigure}
    \caption{Evaluation loss versus scaled optimization steps (left) and CKA similarity heatmap (right). The left panel shows that, across different data budgets, the evaluation loss curves align well after normalizing the optimization steps, indicating that the scaled dynamics are largely insensitive to the choice of learning rate. The right panel exhibits a series of distinct ($5\times5$) high-similarity blocks along the diagonal, where each block corresponds to a fixed data budget. This suggests that models trained with different learning rates but the same data budget learn highly similar internal representations.}
    \label{fig:learning_rate_study}
\end{figure*}

To characterize the best performance attainable under a fixed data budget and model size, we require an evaluation metric that can be monitored continuously throughout training. Following prior work~\cite{scaling_dr}, we use the evaluation loss as our primary measure of model performance. We define the \emph{Minimum Achievable Evaluation Loss} (MAEL) as the minimum evaluation loss attained during training under a fixed training recipe:
\begin{equation}
\mathrm{MAEL}
=
\min_{0 \le t \le T}
L_{\mathrm{eval}}(\theta_t),
\end{equation}
where $\theta_t$ denotes the model parameters after $t$ optimization steps.

A natural question is whether this minimum is sensitive to the learning rate, which strongly influences optimization dynamics. Specifically, we conduct a learning rate sweep on BERT-Medium~\cite{bert} using unique data budgets ranging from 5K to 320K samples and learning rates from $3.125\times10^{-7}$ to $2\times10^{-5}$. All experiments use a constant learning rate without decay.

To compare optimization trajectories across different learning rates, we normalize the training progress by defining the scaled training step as
\begin{equation}
\text{scaled step}=\frac{\text{step}\times\eta}{2\times10^{-5}}.
\end{equation}

\begin{table}[t]
    \centering
    \caption{Learning-rate invariance on BERT-Medium. For each unique-sample budget $U$, we report the mean, standard deviation, and max--min range of MAEL across the five learning rates that reach MAEL: $1.25\times10^{-6}$, $2.5\times10^{-6}$, $5\times10^{-6}$, $1\times10^{-5}$, and $2\times10^{-5}$. Alignment RMSE is computed after interpolating the evaluation-loss curves on their common scaled-step interval.}
    \label{tab:lr_invariance}
    \small
    \begin{tabular}{rcccc}
        \toprule
        $U$ & \#LR & MAEL mean & MAEL std. & MAEL range / RMSE \\
        \midrule
        5K & 5 & 1.108 & 0.015 & 0.039 / 0.026 \\
        10K & 5 & 0.972 & 0.005 & 0.014 / 0.009 \\
        20K & 5 & 0.844 & 0.004 & 0.009 / 0.009 \\
        40K & 5 & 0.729 & 0.009 & 0.021 / 0.008 \\
        80K & 5 & 0.630 & 0.007 & 0.019 / 0.007 \\
        160K & 5 & 0.537 & 0.010 & 0.026 / 0.010 \\
        320K & 5 & 0.463 & 0.006 & 0.016 / 0.011 \\
        \bottomrule
    \end{tabular}
\end{table}

After rescaling, models trained with the same unique data budget exhibit closely aligned learning curves and converge to almost the same minimum evaluation loss. Table~\ref{tab:lr_invariance} quantifies this effect: across data budgets, the average MAEL standard deviation over the five MAEL-reaching learning rates is only 0.008, and the average max--min range is 0.021. By contrast, changing the unique-data budget for BERT-Medium in the main scaling grid changes MAEL by 0.656, and the full 35-point model--data grid spans 1.561 in MAEL. Thus the typical learning-rate-induced MAEL range is only 3.1\% of the BERT-Medium data-scaling range and 1.3\% of the full model--data scaling range. The scaled-step evaluation-loss curves also align closely, with an average interpolation RMSE of 0.011 across budgets.

To further examine whether these models learn similar representations, we compute CKA similarity~\cite{cka} using representations extracted from 1,024 query-document pairs. We compare checkpoints that attain the minimum evaluation loss under different learning rates and observe consistently high CKA scores (close to 1) for models trained with the same unique data budget.

These observations suggest that, in dense retrieval post-training where parameter updates are relatively small and learning rates are low, the learning rate primarily reparameterizes optimization time. Although different learning rates lead to different optimization trajectories, they converge to nearly identical representations and achieve almost the same minimum evaluation loss. We further examine the relationship between MAEL and downstream retrieval performance in Section~\ref{sec:correlation_with_retrieval_metrics}, where we show that MAEL is strongly correlated with standard retrieval metrics.

Therefore, MAEL is substantially less sensitive to the choice of learning rate than the evaluation loss measured at a fixed training step. Throughout this paper, we adopt MAEL as our primary evaluation metric.

\begin{figure*}[t]
    \centering
    \begin{subfigure}{0.45\linewidth}
        \centering
        \includegraphics[width=\linewidth]{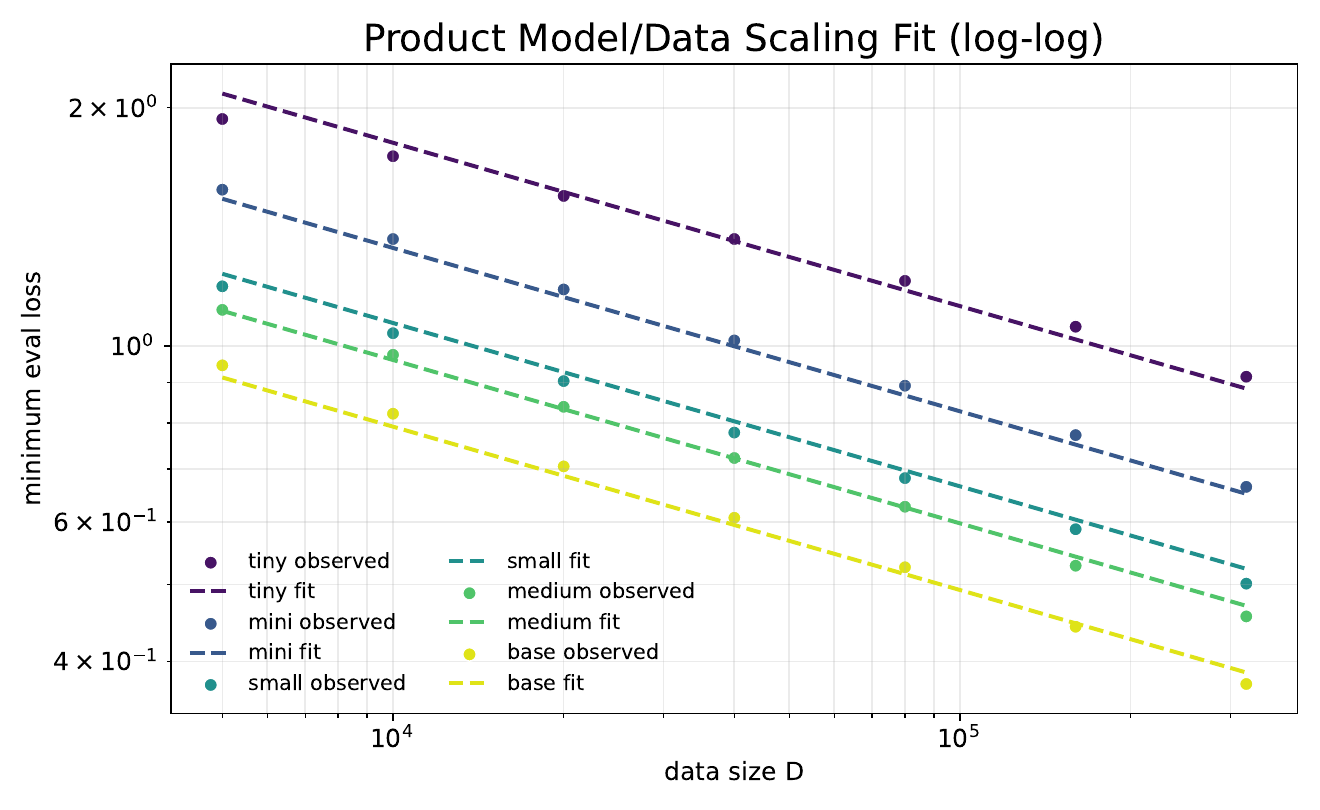}
        \label{fig:eval_loss_data}
    \end{subfigure}
    \begin{subfigure}{0.45\linewidth}
        \centering
        \includegraphics[width=\linewidth]{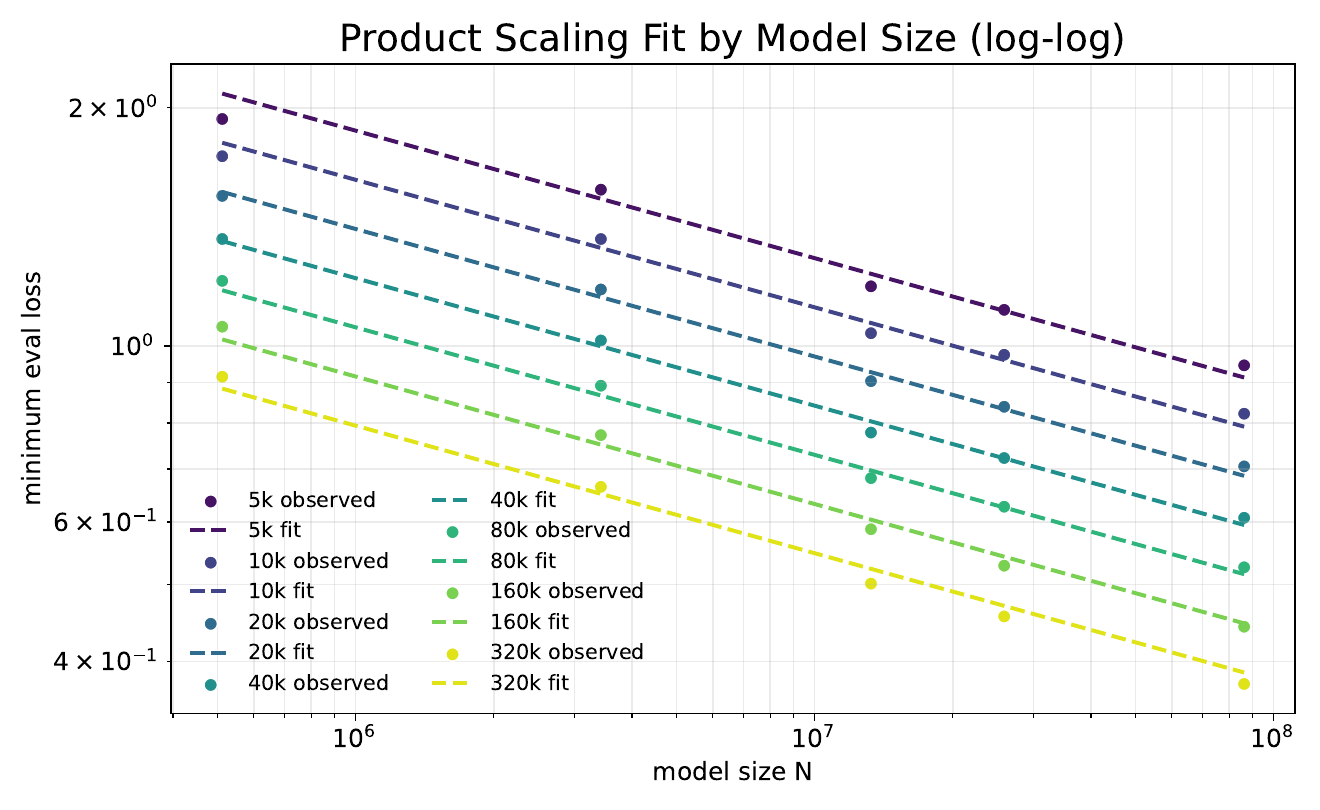}
        \label{fig:eval_loss_param}
    \end{subfigure}
    \caption{MAEL scaling with unique data and model size. The left panel shows MAEL as a function of the number of unique MS MARCO training samples for each BERT model size, while the right panel shows MAEL as a function of non-embedding parameter count for each data budget. The approximately linear and parallel trends in both log--log views motivate the multiplicative model--data scaling law in Eq.~\eqref{eq:multiplicative_style}.}
    \label{fig:eval_vs_model_data}
\end{figure*}

\subsection{Experiment Setup}

We fine-tune a family of BERT models~\cite{bert} on the MS MARCO Passage Ranking dataset~\cite{msmarco} using a contrastive learning objective~\cite{info_nce}. The main scaling grid varies both model size and the number of unique training samples. Model sizes range from BERT-Tiny to BERT-Base, and data budgets range from 5K to 320K unique query--document pairs. We report model scale using non-embedding parameter counts when fitting scaling laws, since embedding parameters are largely fixed by the vocabulary and do not contribute to the same representation-depth scaling. Full model configurations are provided in Table~\ref{tab:model_config}.

To keep the optimization recipe consistent across the scaling grid, all models are trained with AdamW and a constant learning rate, without warmup or learning-rate decay. The learning rate is adjusted only as a function of model size: BERT-Tiny uses $1\times10^{-5}$, and the rate is halved for each successive model variant, yielding $6.25\times10^{-7}$ for BERT-Base. Unless otherwise specified, weight decay is set to zero in the main scaling experiments; the effect of weight decay is studied separately in Section~\ref{sec:weight_decay_scaling}. Training uses four NVIDIA RTX 5090 GPUs with a global batch size of 1,024. For each positive query--document pair, 256 negative passages are randomly sampled during training.

To measure the full optimization trajectory rather than a single fixed-step checkpoint, we periodically evaluate the loss on a fixed evaluation corpus containing 100,000 passages shared across all experiments. Each run is trained for sufficiently many optimization steps to pass the minimum of the evaluation-loss curve and enter the overfitting regime. 

These choices are intended to isolate the effect of model scale and unique-data budget while keeping the training recipe fixed. Appendix~\ref{app:training_details} gives the complete experimental setup, including model architectures, data budgets, optimizer hyperparameters, precision, gradient accumulation, hardware, and evaluation configuration; these details are summarized in Tables~\ref{tab:model_config} and~\ref{tab:training_config}.

\section{Scaling laws for MAEL}
In this section, we study MAEL from four perspectives: its predictability, practical relevance, optimization cost, and improvability. 
In Sec~\ref{sec:model_and_data_scaling}, we establish a joint scaling law for MAEL with respect to model size and unique training data. Sec.~\ref{sec:held_out_and_extrapolation_validation} further examines whether MAEL can be reliably predicted across model sizes and data budgets, while Sec.~\ref{sec:correlation_with_retrieval_metrics} establishes its strong alignment with downstream retrieval performance. Sec.~\ref{sec:training_step_scaling} then studies how much training is required to reach MAEL, and Sec.~\ref{sec:weight_decay_scaling} investigates how weight decay can further reduce MAEL.

\begin{figure*}[t]
    \centering
    \begin{subfigure}{0.3\linewidth}
        \centering
        \includegraphics[width=\linewidth]{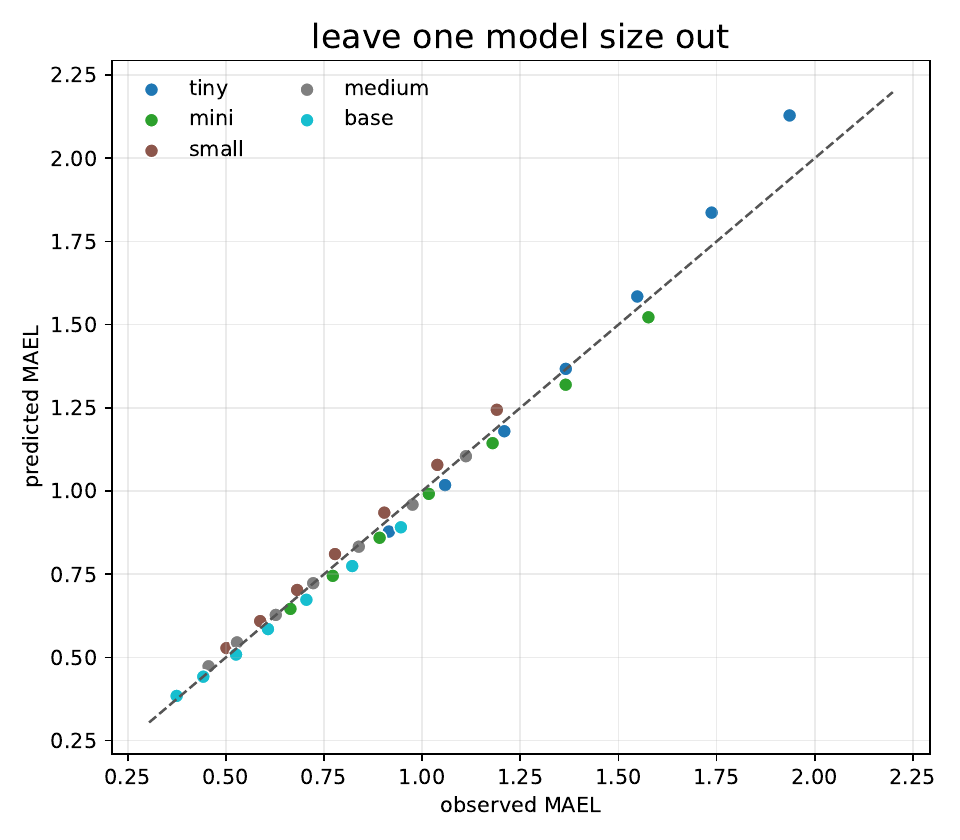}
        \caption{Leave one model size out.}
        \label{fig:heldout_model_size}
    \end{subfigure}
    \begin{subfigure}{0.3\linewidth}
        \centering
        \includegraphics[width=\linewidth]{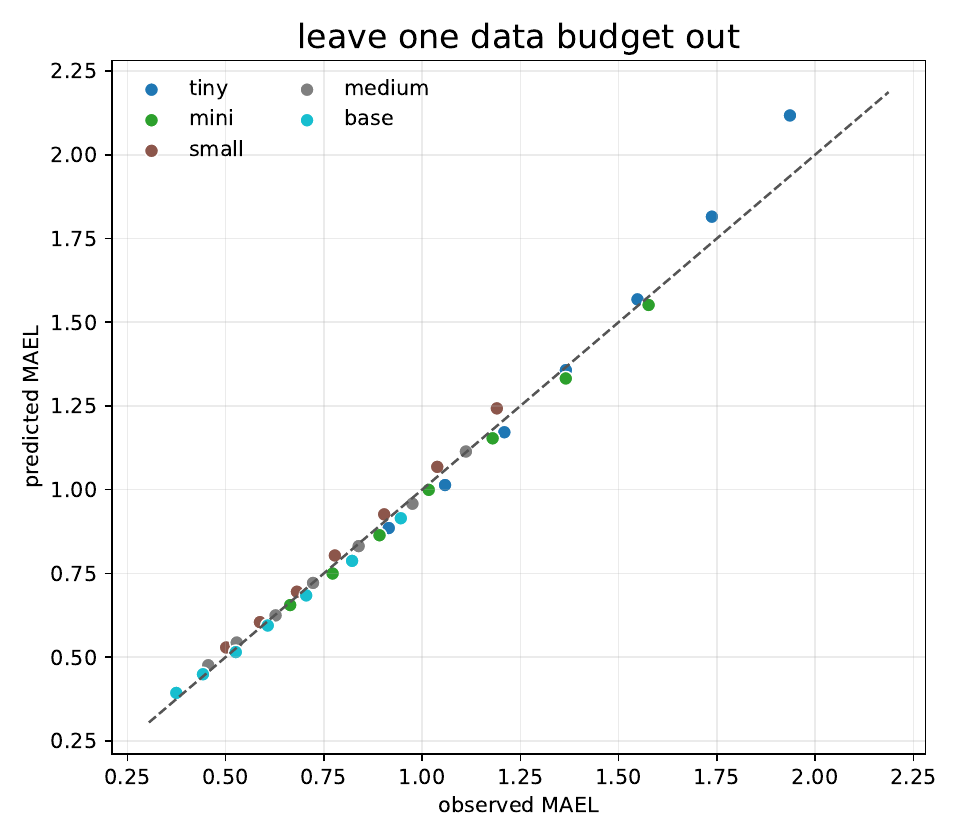}
        \caption{Leave one data budget out.}
        \label{fig:heldout_data_budget}
    \end{subfigure}
    \begin{subfigure}{0.3\linewidth}
        \centering
        \includegraphics[width=\linewidth]{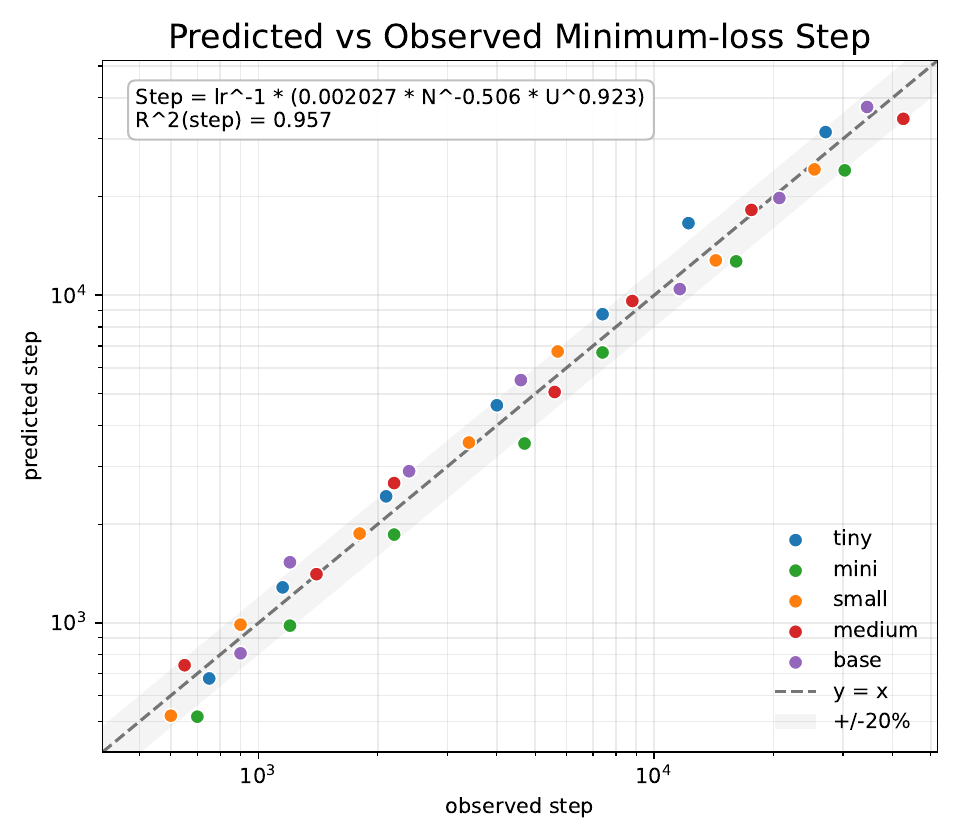}
        \caption{ Optimal training steps}
        \label{fig:step_scaling}
    \end{subfigure}
    \caption{
    (a)--(b) Held-out predictions from the multiplicative MAEL scaling law. Each panel fits Eq.~\eqref{eq:multiplicative_style} without the shown held-out axis value and compares predicted MAEL against observed MAEL. The dashed diagonal indicates perfect prediction. (c) Predicted versus observed optimal training steps. The proposed step scaling law achieves an $R^2$ of 0.95 across different learning rates, model sizes, and unique sample budgets.
    }
    \label{fig:heldout_extrapolation}
\end{figure*}

\subsection{Model and data scaling}
\label{sec:model_and_data_scaling}
We fine-tune the BERT model family on multiple subsets of the MS MARCO training set with varying numbers of unique training samples. Since models trained with different data budgets enter the overfitting regime at different stages of optimization, we adopt a constant learning rate throughout training to ensure that the observed differences arise solely from the available data budget rather than differences in learning rate schedules. During training, we periodically record the evaluation loss and continue optimization until each model reaches its MAEL.

Figure~\ref{fig:eval_vs_model_data} illustrates the relationship between MAEL and both the number of non-embedding parameters and the number of unique training samples. In the log--log domain, MAEL exhibits an approximately linear relationship with both model size and data size across all experimental settings. Notably, the approximately parallel lines observed in both scaling directions indicate that changing the data budget primarily shifts the model scaling curve vertically, while changing model size similarly shifts the data scaling curve. Such behavior is naturally captured by a multiplicative formulation, where improvements from model scaling and data scaling interact proportionally rather than contributing independent additive reductions in loss. This observation is also consistent with recent studies showing that fine-tuning loss can be well described by multiplicative scaling laws~\cite{forgetting_scaling_icml25, forgetting_scaling_arxiv}. Motivated by both this empirical evidence and our observations, we fit MAEL using the following multiplicative scaling law:
\begin{equation}
\mathrm{MAEL}(N,U)
= A\frac{1}{N^{\alpha}}\frac{1}{U^{\beta}}+\delta
\label{eq:multiplicative_style}
\end{equation}
where $N$ denotes the number of non-embedding parameters, $U$ denotes the number of unique fine-tuning samples, and $\delta$ represents the irreducible error.

For comparison, we also fit two scaling law formulations that are widely adopted in the pre-training literature:
\begin{equation}
\mathrm{MAEL}(N,U)
= \left(\frac{A}{N}\right)^{\alpha}+\left(\frac{B}{U}\right)^{\beta}+\delta
\label{eq:chinchilla_style}
\end{equation}
\begin{equation}
\mathrm{MAEL}(N,U)
= \left(\left(\frac{A}{N}\right)^{\frac{\alpha}{\beta}}+\frac{B}{U}\right)^{\beta}+\delta
\label{eq:kaplan_style}
\end{equation}
Table~\ref{tab:scaling_fit} summarizes the joint fitting results. Among the three formulations, the multiplicative scaling law achieves the best fit, with an $R^2$ value of 0.99 and the lowest RMSE, indicating that MAEL can be accurately predicted from the model size and unique data budget within the observed regime. Unlike pre-training scaling laws, which assume that model capacity and data availability introduce independent limitations on the achievable loss, MAEL is measured after repeated optimization over a finite dataset. In this regime, increasing model capacity is more beneficial when sufficient unique data are available, suggesting an interaction between the two factors. This interaction is better captured by the multiplicative formulation.

All three formulations converge to $\delta=0$. One possible interpretation is that, as both the model size and the number of unique training samples increase while optimization proceeds sufficiently long, the model may eventually capture essentially all task-relevant information in the target domain, driving the evaluation loss toward zero. Overall, these results suggest that MAEL follows a predictable joint scaling law over both model size and unique data budget, with a multiplicative interaction providing a substantially better description than existing pre-training scaling formulations.

\begin{table}[htbp]
\centering
\caption{Joint fitting results of different scaling law formulations. ``Kaplan'' refers to Kaplan style scaling law as shown in Eq.~\eqref{eq:kaplan_style}, ``Chinchilla'' refers to Chinchilla style scaling law as shown in Eq.~\eqref{eq:chinchilla_style} and ``Multiplicative'' is shown in Eq.~\eqref{eq:multiplicative_style}.}
\label{tab:scaling_fit}
    \scalebox{0.85}{
    \begin{tabular}{lccccccc}
        \toprule
        Scaling law &
        $A$ &
        $B$ &
        $\alpha$ &
        $\beta$ &
        $\delta$ &
        $R^2$ &
        RMSE \\
        \midrule
        Kaplan~\cite{kaplan_scaling} & $1.12\times10^{6}$ & 0.22 & 7006.25 & 0.53 & 0 & 0.88 & 0.133 \\
        Chinchilla~\cite{chinchilla_scaling} & 54.89 & 20.41 & 0.31 & 0.36 & 0 & 0.96 & 0.076 \\
        Multiplicative & 100.28 & -- & 0.16 & 0.21 & 0 & \textbf{0.99} & \textbf{0.035} \\
        \bottomrule
    \end{tabular}
    }
\end{table}

\subsection{Held-out and extrapolation validation}
\label{sec:held_out_and_extrapolation_validation}

The high in-sample fit in Table~\ref{tab:scaling_fit} shows that the multiplicative law describes the observed grid, but it does not by itself establish that the law can predict unobserved model scales or data budgets. We therefore evaluate Eq.~\eqref{eq:multiplicative_style} under held-out validation protocols. In all cases, we use the same MAEL target as in the main scaling experiment, use non-embedding parameter counts for $N$, matching the fitted setting in Table~\ref{tab:scaling_fit}.

We consider four validation settings. First, leave-one-model-size-out validation removes all seven data-budget points for one BERT scale, fits the law on the remaining four model sizes, and predicts the held-out model size. Second, leave-one-data-budget-out validation removes all five model-size points for one unique-sample budget, fits on the remaining six budgets, and predicts the held-out budget. Third, small-to-large corner extrapolation fits only on the smaller part of the grid, using Tiny, Mini, and Small models with at most 80K unique samples, and predicts the largest unseen corner consisting of Medium and Base models at 160K and 320K samples. Finally, we also report a broader small-grid extrapolation setting that uses the same 15 small-grid training points and predicts all remaining 20 points outside that grid.

\begin{table}[t]
\centering
\caption{Held-out and extrapolation validation of the multiplicative MAEL scaling law. Relative errors are absolute relative prediction errors in MAEL.}
\label{tab:heldout_extrapolation}
\resizebox{\columnwidth}{!}{
\begin{tabular}{lcccc}
\toprule
Validation & Held-out points & RMSE & Mean rel. err. & Max rel. err. \\
\midrule
Leave one model size out & 35 & 0.0471 & 3.28\% & 9.95\% \\
Leave one data budget out & 35 & 0.0407 & 2.77\% & 9.36\% \\
Small-to-large corner & 4 & 0.0207 & 4.45\% & 5.67\% \\
Small-grid extrapolation & 20 & 0.0264 & 3.14\% & 7.05\% \\
\bottomrule
\end{tabular}
}
\end{table}

Table~\ref{tab:heldout_extrapolation} summarizes the held-out prediction errors, and Figure~\ref{fig:heldout_extrapolation} visualizes representative predicted-versus-observed comparisons. The multiplicative law remains accurate when entire model scales are excluded from fitting, achieving an RMSE of 0.0471 and a mean absolute relative error of 3.28\%. Holding out entire data budgets gives a similar result, with an RMSE of 0.0407 and a mean absolute relative error of 2.77\%. These two protocols are stricter than randomly holding out individual grid points because each test fold removes an entire axis value, forcing the law to recover either an unseen model scale or an unseen data budget from the remaining grid.

The small-to-large tests further probe extrapolation beyond the small-scale fitting region. When fit only on Tiny, Mini, and Small models trained with at most 80K unique samples, the law predicts the Medium/Base and 160K/320K corner with an RMSE of 0.0207 and a mean absolute relative error of 4.45\%. Across all 20 points outside the same small-grid training region, the mean absolute relative error is 3.14\%. The errors remain modest relative to the full observed MAEL range, which spans from 0.375 to 1.936 in these experiments.

These held-out results support the use of the multiplicative law as a predictive scaling relationship rather than only an in-sample curve fit. In particular, the law generalizes across both scaling axes and maintains useful accuracy when extrapolating from smaller models and smaller data budgets to larger configurations.

\subsection{Correlation with retrieval metrics}
\label{sec:correlation_with_retrieval_metrics}

Although MAEL provides a dense and continuously observable target for scaling-law fitting, it is useful only if it reflects the ranking quality that retrieval systems ultimately care about. We therefore directly validate the relationship between MAEL and standard retrieval metrics on the same 35 configurations used in the scaling grid. For each model-size and unique-data-budget pair, we take the checkpoint that attains MAEL and evaluate it on the MS MARCO in-domain validation setting with a 100,000-document retrieval pool and 6,980 validation queries.

\begin{figure}[htbp]
    \centering
    \includegraphics[width=\linewidth]{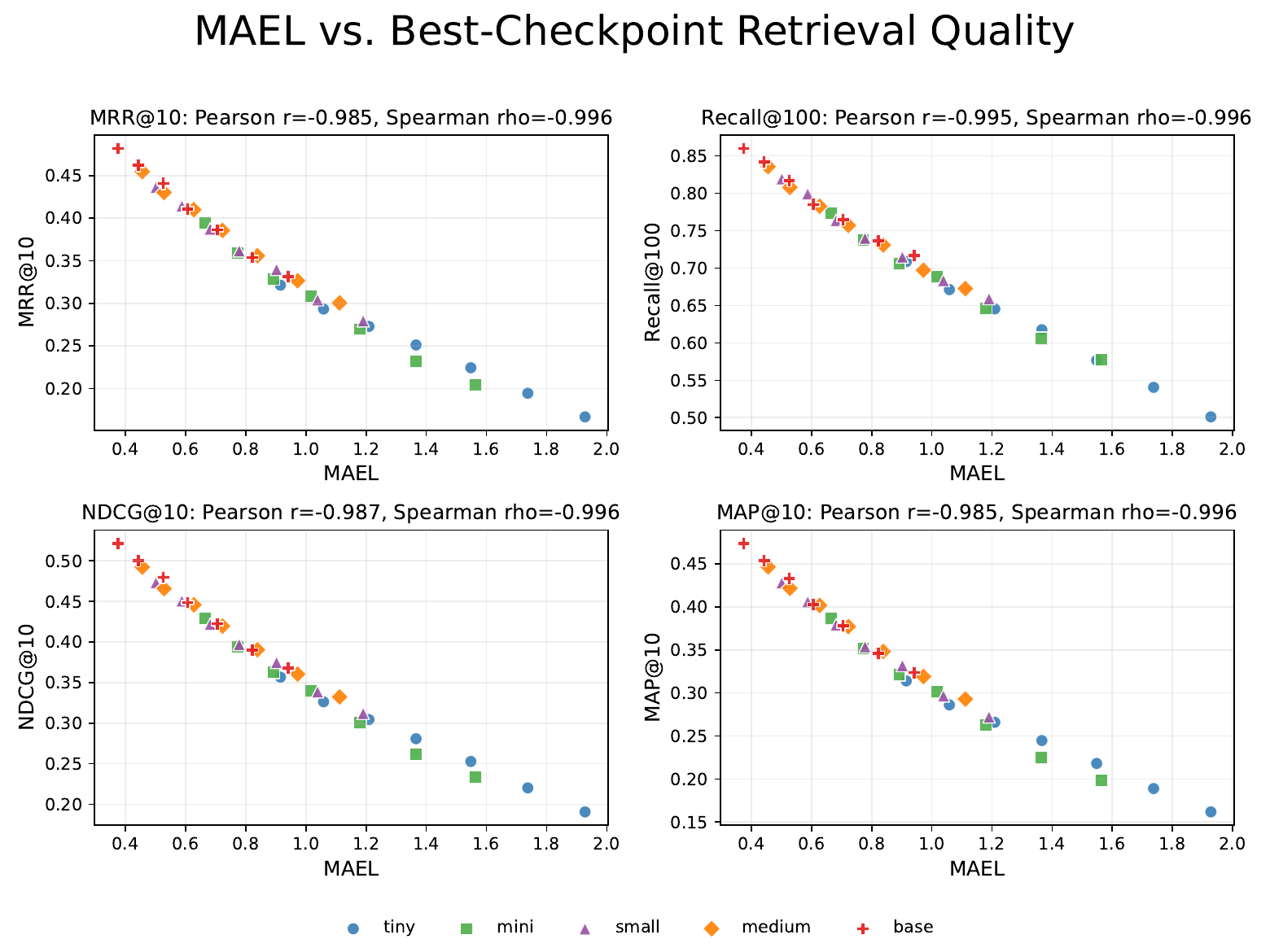}
    \caption{Relationship between MAEL and downstream retrieval metrics across the 35 model-size and unique-data-budget configurations. Lower MAEL consistently corresponds to higher MRR@10, Recall@100, nDCG@10, and MAP@10.}
    \Description{Four scatter plots comparing MAEL against MRR@10, Recall@100, nDCG@10, and MAP@10. Each plot shows a strong negative relationship, indicating that lower MAEL corresponds to better retrieval quality.}
    \label{fig:mael_retrieval_correlation}
\end{figure}

\begin{table}[t]
\centering
\caption{Correlation between MAEL and MS MARCO in-domain retrieval metrics across 35 configurations. Since MAEL is lower-is-better and retrieval metrics are higher-is-better, stronger retrieval quality should appear as a negative correlation.}
\label{tab:mael_retrieval_correlation}
\resizebox{\columnwidth}{!}{
\begin{tabular}{lcccc}
\toprule
Metric & Pearson $r$ & Pearson $p$ & Spearman $\rho$ & Spearman $p$ \\
\midrule
MRR@10 & -0.9853 & $6.44{\times}10^{-27}$ & -0.9958 & $7.58{\times}10^{-36}$ \\
Recall@100 & -0.9948 & $2.59{\times}10^{-34}$ & -0.9958 & $7.58{\times}10^{-36}$ \\
nDCG@10 & -0.9868 & $1.11{\times}10^{-27}$ & -0.9955 & $2.19{\times}10^{-35}$ \\
MAP@10 & -0.9848 & $1.17{\times}10^{-26}$ & -0.9958 & $7.58{\times}10^{-36}$ \\
\bottomrule
\end{tabular}
}
\end{table}

Figure~\ref{fig:mael_retrieval_correlation} and Table~\ref{tab:mael_retrieval_correlation} show that MAEL is strongly and consistently anti-correlated with retrieval effectiveness. The Pearson correlations range from $-0.9848$ to $-0.9948$, and the Spearman rank correlations range from $-0.9955$ to $-0.9958$ across all four retrieval metrics. This indicates that the MAEL ordering nearly matches the ordering induced by true retrieval quality: configurations with lower attainable evaluation loss also retrieve relevant documents more effectively. These results support using MAEL as the primary target for scaling-law analysis rather than merely as a convenient optimization proxy. 

\subsection{Training step scaling}
\label{sec:training_step_scaling}
While the MAEL scaling law determines how low the evaluation loss can be, it does not tell us how long a model must be trained to reach that loss. We therefore next derive a scaling law for the optimal training step. Our previous experiments show that the optimal training steps are approximately inversely proportional to the learning rate after time reparameterization. Furthermore, we observed that increasing model size reduces the required optimization steps and increasing the number of unique samples increases the required optimization steps. Based on these observations, we model the following scaling law for the optimal training steps:
\begin{equation}
\mathrm{Step}_{\mathrm{opt.}}(N,U)
= \frac{1}{\eta}\times\left(KN^{\gamma}U^{\kappa}+\delta'\right)
\label{eq:step_scaling}
\end{equation}
where $N$ denotes the number of model parameters, $U$ is the number of unique training samples, and $\eta$ is the learning rate. Figure~\ref{fig:step_scaling} compares the predicted and observed optimal training steps. The proposed scaling law achieves an $R^2$ of 0.95, indicating that it accurately captures the dependence of the optimal training steps on learning rate, model size, and unique training sample budget.

\subsection{Weight decay scaling}
\label{sec:weight_decay_scaling}
\begin{figure*}[t]
    \centering
    \includegraphics[width=0.9\linewidth]{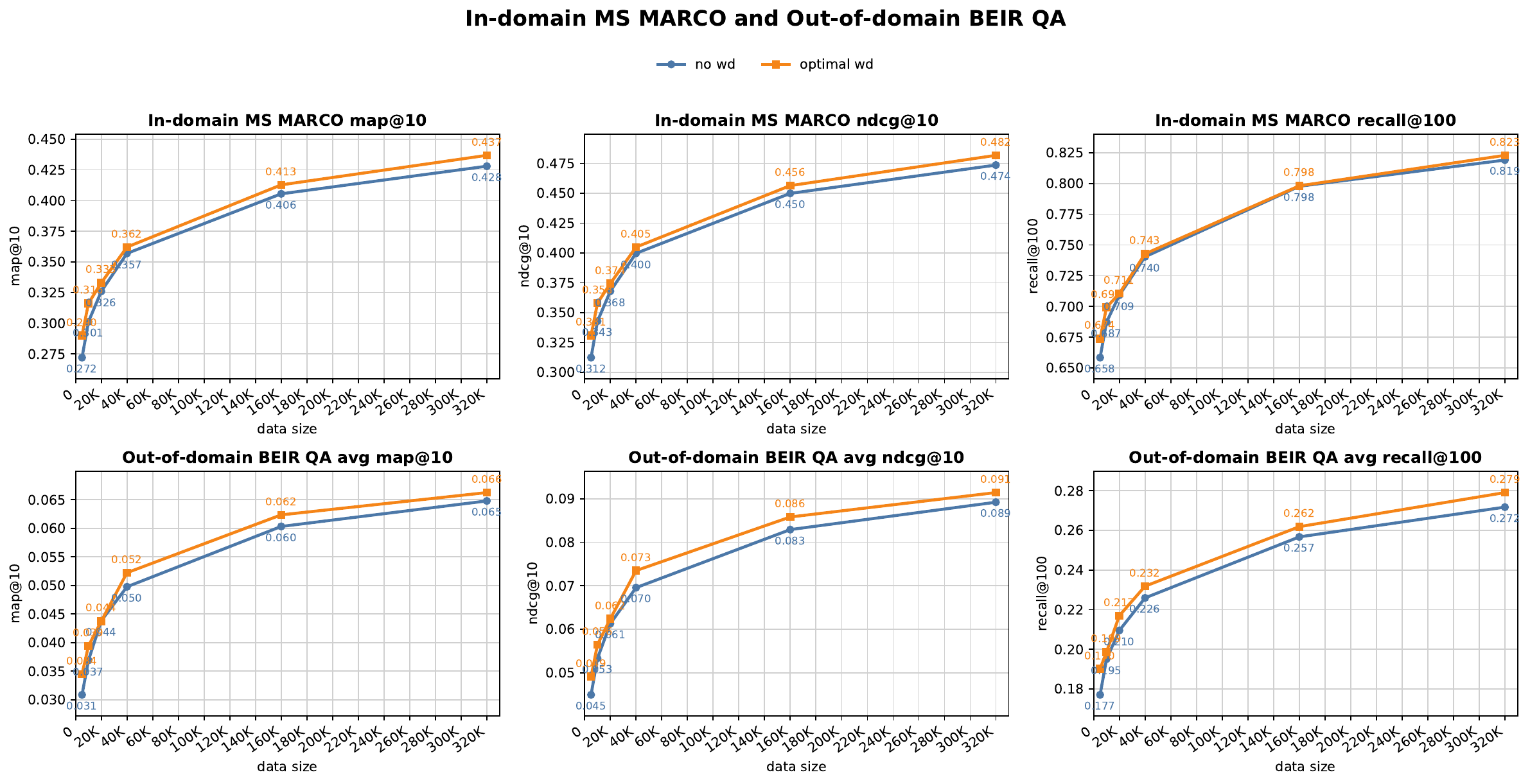}
    \caption{Effect of optimal weight decay on downstream retrieval performance. We compare BERT-Small checkpoints trained with no weight decay and with the scaled optimal weight decay across unique-sample budgets. The top row reports in-domain MS MARCO performance, and the bottom row reports macro-averaged out-of-domain BEIR QA performance over NQ, HotpotQA, and FiQA; columns show MAP@10, nDCG@10, and Recall@100.}
    \label{fig:wd_retrieval}
\end{figure*}
\begin{figure}[htbp]
    \centering
    \includegraphics[width=\linewidth]{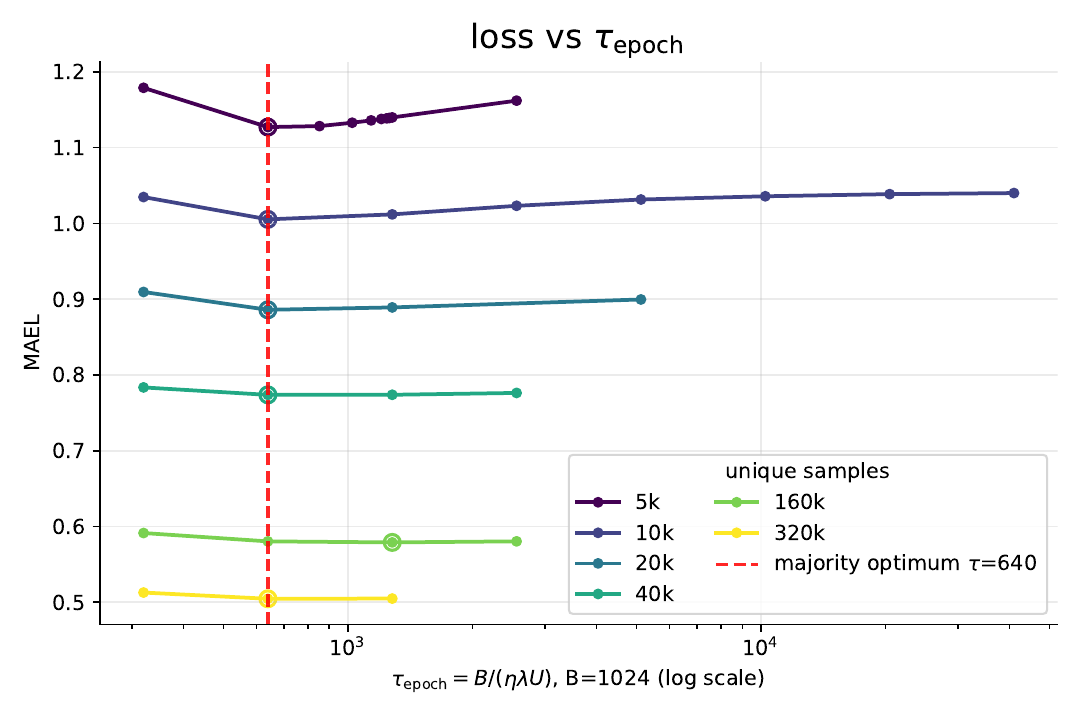}
    \caption{MAEL as a function of the AdamW EMA timescale $\tau_{\mathrm{epoch}}$ for BERT-Small under different unique-sample budgets. Each curve corresponds to a weight-decay sweep after converting the configured weight decay to $\tau_{\mathrm{epoch}}=B/(\eta\lambda U)$. Hollow markers indicate the minimum MAEL on each curve, and the dashed vertical line marks the majority optimal timescale across data budgets.}
    \label{fig:optimal_tau}
\end{figure}
Weight decay is a regularization technique that mitigates model overfitting. To study how the optimal setting changes with the amount of repeated data, we sweep the weight decay coefficient for BERT-Small under different unique sample budgets and define the value that achieves the lowest MAEL as the optimal weight decay. All experiments are conducted using the AdamW optimizer, whose decoupled weight decay formulation follows the original implementation~\cite{adamw}:
\begin{equation}
w_{t}
= (1-\eta_{t}\lambda)w_{t-1}-\eta_{t}\alpha\frac{\hat{m}_t}{\sqrt{\hat{v}_t}+\varepsilon}
\label{eq:adamw_ori}
\end{equation}
where $\eta_t$ denotes the schedule multiplier, which is typically used for learning rate warmup or decay. $\lambda$ is the weight decay coefficient that penalizes large parameter magnitudes, and $\alpha$ is the learning rate. In this formulation, weight decay is decoupled from the gradient-based parameter update, allowing the learning rate and weight decay to be tuned independently. However, modern implementations, including the one used in our experiments, typically adopt the AdamW implementation provided by PyTorch's \texttt{torch.optim} package:
\begin{equation}
\eta_{t}'
= \alpha\eta_{t}
\end{equation}
resulting in the update rule
\begin{equation}
w_{t}
= (1-\eta_{t}'\lambda)w_{t-1}-\eta_{t}'\frac{\hat{m}_t}{\sqrt{\hat{v}_t}+\varepsilon}
\label{eq:adamw_used}
\end{equation}
Consequently, the effective weight decay strength is proportional to $\eta_t'\lambda$. Consequently, when a smaller learning rate is used (e.g., $10^{-6}$ in our experiments), the optimal configured weight decay becomes correspondingly larger, often reaching values on the order of $10^2$. Following previous work, the optimization dynamics can be interpreted through an equivalent exponential moving average (EMA) timescale:
\begin{equation}
\tau_{\mathrm{epoch}}
= \frac{B}{\eta\lambda U}
\label{eq:adamw_ema}
\end{equation}
where $\eta$ is the learning rate, $\lambda$ is the configured weight decay in the PyTorch implementation, $U$ denotes the number of unique training samples, and $B$ is the batch size. 

Figure~\ref{fig:optimal_tau} shows the relationship between MAEL and $\tau_{\mathrm{epoch}}$ under different unique sample budgets. In terms of the raw configured weight decay, the optimum shifts as $U$ changes because the effective shrinkage per epoch depends on $\eta\lambda U$. After converting each sweep to the EMA timescale in Eq.~(\ref{eq:adamw_ema}), however, the minima from different data budgets concentrate around a similar $\tau_{\mathrm{epoch}}$.

These results suggest that, for a fixed model and learning-rate setting, the optimal weight decay during dense-retrieval fine-tuning is better described by an approximately invariant EMA timescale than by a fixed configured coefficient. Equivalently, when other hyperparameters are fixed, $\eta\lambda U$ should remain approximately constant as the unique-data budget changes.

Finally, we investigate whether the lower MAEL obtained with the optimal weight decay translates into improved retrieval performance. We train two sets of BERT-Small models under different unique sample budgets: one without weight decay ($\lambda=0$), and one using the optimal weight decay identified by our scaling law. Specifically, the model trained on the 320K unique sample budget uses a weight decay of 2, while the weight decay is doubled whenever the unique sample budget is halved, with all other hyperparameters kept identical to the main experiments.

We evaluate the checkpoints achieving MAEL on four retrieval benchmarks using Recall@100, MAP@10, and nDCG@10. The evaluation includes the in-domain MSMARCO test set and three out-of-domain benchmarks widely used for open-domain question answering: NQ~\cite{nq}, HotpotQA~\cite{hotpot_qa}, and FiQA~\cite{fiqa}.

Figure~\ref{fig:wd_retrieval} compares retrieval performance with and without the optimal weight decay across different unique sample budgets. Consistently across both in-domain and out-of-domain datasets, the lower MAEL achieved by the optimal weight decay leads to improved retrieval effectiveness, demonstrating that optimizing MAEL also translates into better downstream retrieval performance.

\section{Compute Allocation under Scaling Laws}

The scaling laws above allow us to ask a practical allocation question: under a fixed fine-tuning compute budget, is it more efficient to spend additional resources on a larger pretrained model or on more unique in-domain data? This question is especially relevant in post-training, where the pre-training cost of a larger checkpoint is usually not charged to the downstream fine-tuning budget.

The compute required to reach MAEL is determined by the number of optimal training steps in Eq.~\eqref{eq:step_scaling}. Following the standard approximation used in scaling-law analyses that fine-tuning compute is proportional to the product of model size and training steps (approximately 6N FLOPs per training token for Transformer training) ~\cite{kaplan_scaling}, and omitting additive constants while absorbing the learning-rate-dependent factor into a recipe-specific constant, the fitted step scaling gives
\begin{equation}
\mathrm{Step}_{\mathrm{opt.}}(N,U)
\propto
N^{-0.5064}U^{0.9231}.
\end{equation}
Since fine-tuning compute is proportional to model size times training steps, we obtain
\begin{equation}
C_{\mathrm{ft}}
\propto
6N\mathrm{Step}_{\mathrm{opt.}}
\propto
N^{0.4936}U^{0.9231}.
\label{eq:finetune_compute_scaling}
\end{equation}

Under the multiplicative MAEL law in Eq.~\eqref{eq:multiplicative_style}, a fixed-compute frontier does not generally produce a finite interior optimum. More generally, suppose
\begin{equation}
L(N,U)=AN^{-\alpha}U^{-\beta},
\qquad
C=QN^{a}U^{b}.
\end{equation}
Substituting $U=(C/(QN^{a}))^{1/b}$ gives
\begin{equation}
L(N)
=
A\left(\frac{C}{Q}\right)^{-\frac{\beta}{b}}
N^{-\alpha+\frac{a\beta}{b}}.
\end{equation}
Therefore, unless $\alpha/a=\beta/b$, the loss is monotonic along the fixed-compute frontier rather than minimized at an interior allocation. The fitted multiplicative law gives
\begin{equation}
\mathrm{MAEL}(N,U)
=
100.280161\,N^{-0.161017}U^{-0.206251}.
\end{equation}
Combining it with Eq.~\eqref{eq:finetune_compute_scaling}, the compute efficiency of model scaling is $\alpha/a=0.161017/0.4936=0.3262$, while that of unique-data scaling is $\beta/b=0.206251/0.9231=0.2234$. Thus, the multiplicative law predicts that increasing the pretrained model size is approximately $0.3262/0.2234\approx1.46$ times as fine-tuning-compute-efficient as increasing the unique in-domain data budget for reducing MAEL.

To test whether this prescription is merely an artifact of the multiplicative formulation, we instantiate a counterfactual allocation using the additive Chinchilla-style law in Eq.~\eqref{eq:chinchilla_style}, which also fits the observations reasonably well ($R^2=0.96$). For the additive law
\begin{equation}
L_{\mathrm{add}}(N,U)=AN^{-\alpha}+BU^{-\beta},
\qquad
C_{\mathrm{ft}}=QN^{a}U^{b},
\end{equation}
the fixed-compute optimum satisfies
\begin{equation}
\alpha AN^{-\alpha}
=
\frac{a\beta}{b}BU^{-\beta},
\end{equation}
which gives the compute-optimal data budget for a chosen model size:
\begin{equation}
U_{\mathrm{add}}^{*}(N)
=
\left(\frac{a\beta B}{b\alpha A}\right)^{\frac{1}{\beta}}
N^{\frac{\alpha}{\beta}}.
\label{eq:additive_compute_optimal_data}
\end{equation}
Using the fitted additive parameters $A=54.89$, $B=20.41$, $\alpha=0.31$, and $\beta=0.36$, together with $a=0.4936$ and $b=0.9231$, Eq.~\eqref{eq:additive_compute_optimal_data} becomes
\begin{equation}
U_{\mathrm{add}}^{*}(N)
\approx
0.0170\,N^{0.861}.
\end{equation}
All quantities use the same raw units as the scaling-law fit: $N$ is the number of non-embedding parameters and $U$ is the number of unique fine-tuning samples.

We use BERT-Medium as the reference model. With $N_{\mathrm{med}}=2.59\times10^{7}$, the additive law recommends
\begin{equation}
U_{\mathrm{med}}^{*}
\approx
4.12\times10^{4}
\end{equation}
unique samples, which lies close to our 40K experimental grid point. We then keep the same fine-tuning compute budget and ask how much data BERT-Base should use. With $N_{\mathrm{base}}=8.62\times10^{7}$,
\begin{equation}
U_{\mathrm{base}}
=
\left(
\frac{N_{\mathrm{med}}^{0.4936}(U_{\mathrm{med}}^{*})^{0.9231}}
{N_{\mathrm{base}}^{0.4936}}
\right)^{\frac{1}{0.9231}}
\approx
2.17\times10^{4}.
\end{equation}
Thus, the additive law defines a compute-optimal BERT-Medium allocation near 41.2K samples, while the same compute budget corresponds to a deliberately more model-heavy BERT-Base allocation near 21.7K samples.

\begin{table}[htbp]
\centering
\caption{Comparison between the additive compute-optimal allocation and a larger-model allocation under a fixed compute budget. RC. denotes relative compute. Predicted results are obtained from the scaling laws, while observed results correspond to the nearest available experimental configurations.}
\label{tab:compute_efficiency_case_study}
\scalebox{0.8}{
\begin{tabular}{lcccccc}
\toprule
Allocation & Model & $N$ & $U$ & Step & RC. & MAEL \\
\midrule
Chinchilla & BERT-Medium & 25.88M & 41.2K & -- & 1.00 & 0.717 (pred.) \\
Multiplicative & BERT-Base & 86.24M & 21.7K & -- & 1.00 & \textbf{0.675} (pred.) \\
\midrule
Chinchilla & BERT-Medium & 25.88M & 41.2K & 4k & 1.00 & 0.720 (obs.) \\
Multiplicative & BERT-Base & 86.24M & 21.7K & 1.3k & 1.08 & \textbf{0.690} (obs.) \\
\bottomrule
\end{tabular}
}
\end{table}

For the observed comparison, we compute the realized fine-tuning cost directly from the experimental optimal steps rather than from the fitted step law:
\begin{equation}
C_{\mathrm{obs}}=6N\mathrm{Step}_{\mathrm{obs}}.
\end{equation}
The BERT-Medium setting reaches its observed minimum at 4000 steps, giving
\begin{equation}
C_{\mathrm{med}}
=
6\times 25.88\mathrm{M}\times 4000
=
6.21\times10^{11}.
\end{equation}
The BERT-Base setting reaches its observed minimum at 1300 steps, giving
\begin{equation}
C_{\mathrm{base}}
=
6\times 86.24\mathrm{M}\times 1300
=
6.73\times10^{11}.
\end{equation}
Thus the observed BERT-Base run uses
\begin{equation}
\frac{C_{\mathrm{base}}}{C_{\mathrm{med}}}
=
\frac{86.24\mathrm{M}\times1300}{25.88\mathrm{M}\times4000}
\approx
1.08
\end{equation}
times the compute of the BERT-Medium run. Equivalently, matching the BERT-Base compute exactly would allow the BERT-Medium model to train for
\begin{equation}
\mathrm{Step}_{\mathrm{med}}^{\mathrm{equal}}
=
\frac{86.24\mathrm{M}\times1300}{25.88\mathrm{M}}
\approx
4331
\end{equation}
steps. Therefore, the two observed settings are close to compute-matched: BERT-Base uses only 8\% more realized fine-tuning compute while achieving a lower MAEL, 0.690 compared with 0.720. This empirical comparison supports the multiplicative-law prescription that, in this regime, allocating compute toward a larger pretrained model can be more effective than allocating it toward additional unique data.

These results provide empirical support for the compute-efficiency guidance implied by the multiplicative MAEL law. In the observed regime, when a larger compatible pretrained checkpoint can be trained and served, allocating fine-tuning compute toward the larger model is more effective than spending the same budget on additional unique in-domain data. Once model size is constrained by hardware, checkpoint availability, or deployment latency, increasing the amount of unique in-domain data remains the next effective way to reduce MAEL.

\section{Conclusion}
This work studies scaling laws for dense retrieval under data-constrained fine-tuning, where supervised retrieval data is limited and repeatedly reused. We introduce minimum achievable evaluation loss (MAEL) as a stable measure of the best performance attainable under fixed model-size and unique-data budgets, and show that MAEL follows a simple scaling law with respect to model size and data budget across 35 BERT-family dense retrievers fine-tuned on MS MARCO subsets. Beyond serving as a scaling target, MAEL strongly correlates with downstream retrieval effectiveness across standard retrieval metrics. We further demonstrate that MAEL convergence and weight-decay selection can be predicted from training dynamics, enabling more efficient fine-tuning under limited data. Our findings provide practical guidance for allocating fine-tuning resources: scaling pretrained model size is often more compute-efficient than collecting additional in-domain data in the studied regime, while increasing unique data remains an effective strategy when model capacity is constrained. More broadly, this work extends scaling-law analysis beyond pre-training to data-scarce post-training scenarios.

\appendix

\section{Training details}
\label{app:training_details}

\begin{table}[htbp]
\centering
\caption{BERT model configurations used in our experiments. Values in parentheses denote non-embedding parameters, excluding the token and position embedding layers.}
\label{tab:model_config}
\begin{tabular}{lccc}
\toprule
Model & Layers & Hidden Size & Params \\
\midrule
BERT-Tiny   & 2  & 128 & 4.48M (0.51M)  \\
BERT-Mini   & 4  & 256 & 11.37M (3.42M) \\
BERT-Small  & 4  & 512 & 29.16M (13.27M) \\
BERT-Medium & 8  & 512 & 41.77M (25.88M) \\
BERT-Base   & 12 & 768 & 110.07M (86.24M) \\
\bottomrule
\end{tabular}
\end{table}

\begin{table}[htbp]
\small
\centering
\caption{Training and evaluation configuration shared across all experiments unless otherwise specified.}
\label{tab:training_config}

\begin{tabularx}{\columnwidth}{>{\bfseries}lX}
\toprule
Item & Value \\
\midrule
Dataset &
MS MARCO Passage Ranking \\

Training sample budgets &
5K, 10K, 20K, 40K, 80K, 160K, and 320K unique query--document pairs. \\

Loss function & infoNCE
\\

Negative sampling &
256 randomly sampled negative passages for each positive query--document pair. \\

Evaluation corpus &
A fixed corpus of 100,000 passages shared across all experiments. \\

Optimizer &
AdamW \\

Learning rate &
$1\times10^{-5}$ for BERT-Tiny, halved for each successive model variant down to $6.25\times10^{-7}$ for BERT-Base. \\

Learning rate schedule & None (constant learning rate). \\

Warmup &
0 steps. \\

AdamW $\beta$ &
$(\beta_1,\beta_2)=(0.9, 0.999)$ \\

AdamW $\epsilon$ & $1\times10^{-8}$ \\

Weight decay & 0 \\

Mixed precision & BF16 \\

Global batch size & 1,024 \\

Hardware & 4 $\times$ NVIDIA RTX 5090 GPUs. \\

\bottomrule
\end{tabularx}
\end{table}

\clearpage
\bibliographystyle{ACM-Reference-Format}
\bibliography{main}

\end{document}